\documentclass[a4paper,11pt]{article}

\pdfoutput=1 

\usepackage{./style_files/jheppub} 

\usepackage[T1]{fontenc} 

\title{\boldmath Fisher Information of a Black Hole Spacetime}

\author[a,1]{Everett Patterson,\note{Corresponding author.}}
\author[a,b]{Robert B.  Mann}

\affiliation[a]{Department of Physics and Astronomy, University of Waterloo, \\ Waterloo, Ontario, Canada N2L 3G1}
\affiliation[b]{Perimeter Institute for Theoretical Physics, \\ 31 Caroline St N, Waterloo, Ontario, Canada N2L 2Y5}

\emailAdd{ea2patte@uwaterloo.ca}
\emailAdd{rbmann@uwaterloo.ca}

\abstract{ Relativistic quantum metrology is the study of optimal measurement procedures within systems that have both quantum and relativistic components. Here we use Unruh-DeWitt detectors coupled to a massless scalar field as probes of thermal parameters in different spacetimes via a relativistic quantum metrology analysis. We   consider both $(2+1)$-dimensional anti-de Sitter and BTZ black hole spacetimes. We compute the Fisher information to  identify characteristics of the black hole spacetime and to compare it to a uniformly accelerating detector in anti-de Sitter space. We find the dependence of the Fisher information  on temperature, detector energy gap,  black hole mass, interaction time, and the initial state of the detector.  
We  identify  strategies that maximize the Fisher information and therefore the precision of estimation.}

\usepackage{braket}
\usepackage{color}
\usepackage[dvipsnames]{xcolor}

\newcommand{\D}[2]{\frac{\partial #1}{\partial #2}}

\renewcommand{\vec}[1]{\boldsymbol{#1}}
\newcommand{\F}{\mathcal{F}}

\newcommand{\I}{\mathcal{I}}

\newcommand{\be}{\begin{equation}}
\newcommand{\ee}{\end{equation}}
\newcommand{\ba}{\begin{eqnarray}}
\newcommand{\ea}{\end{eqnarray}}
\newcommand{\nn}{\nonumber}
\usepackage{colonequals}
\newcommand{\ce}{\colonequals}
\newcommand{\rh}{r_{\rm h}}

\usepackage{ulem}

\usepackage[utf8]{inputenc}
\usepackage[T1]{fontenc}
\usepackage{fancyhdr}
\usepackage{lipsum}
\usepackage{graphicx}
\usepackage{titlesec}
\usepackage{appendix}
\usepackage[sectionbib]{chapterbib}
\usepackage[breakwords]{truncate}
\usepackage{lastpage}
\usepackage{amsmath}
\usepackage{hyperref}
\usepackage{amssymb}
\usepackage[font={small}]{caption}
\usepackage{physics}

\begin{document}
\maketitle
\flushbottom

\newpage

\section{Introduction}

For much of the past century, a great number of theorists have made it their quest to reconcile the theories of quantum mechanics and general relativity. While these two theories are independently backed by some of the most robust experimental evidence within their respective regimes of validity, efforts to unite them into an overarching theory of quantum gravity has thus far remained elusive.


A recent approach toward understanding quantum gravity has been that of relativistic quantum information \cite{MannRalph2012RQI}, which seeks to consider both the effects of relativity on quantum information protocols and  how quantum information tools can help us describe and probe relativistic systems. A particular example is that of the pragmatically motivated relativistic quantum metrology \cite{ABSAF2014RQM}, whose goal is to optimize the measurement procedure for physical parameters in systems that have both quantum and relativistic components. 

While quantum metrology has been around for some time now \cite{GLM2011,BraunsteinCaves1994,Paris2009qe4qt}, the consideration of relativistic influences is a more recent development. Its importance in gravitational physics began with its use in enhancing the detection of gravitational waves using LIGO/VIRGO \cite{LIGO2011}.
There have subsequently been other  improvements in gravitional wave detection \cite{GDDSSV2013} by exploiting quantum squeezed states in interferometers \cite{Caves1981}.  

Interest in quantum metrology is becoming increasingly intertwined with relativistic ideas. Recent examples include the investigation of quantum metrology with indefinite causal order \cite{ZYC2020}, the use of quantum clocks to improve the measurement of gravitational time dilation \cite{CGP2022}, the interaction of Bose-Einstein condensates with gravitational waves \cite{Schutzhold2018,RobbinsAfshordiMann2019}, as well as estimations of the Unruh-Hawking effect \cite{AAF2010} and of spacetime parameters for a Schwarzchild model of the Earth \cite{BDURF2014}, for the Robertson-Walker expanding universe \cite{WTJF2015}, and for an expanding de Sitter spacetime \cite{HFZF2018}.

In this paper we employ the tools and techniques of relativistic quantum metrology, notably Fisher information and the Unruh-DeWitt (UDW) detector model \cite{Unruh1976,DeWitt1979}, to the task of discriminating between spacetimes. Fisher information is a key figure of merit in estimating parameters that do not (at least directly) correspond to quantum observables.  The idea is to determine the sensitivity of the state of a system with respect to changes in such parameters, thereby providing information about the maximal precision that can be possibly achieved in a metrological task, via the Cram\'er-Rao bound \cite{Rao1992,Cramer1999}. 

In particular, we shall consider how Fisher information can be used to probe the semiclassical properties of black holes.  Previous analyses of de Sitter and anti-de Sitter spacetimes have been performed in $(3+1)$ dimensions \cite{DuMann2021}, and an extension to black holes is a natural progression.
In general this is a rather challenging step, since the computation of detector response involves a sum over field modes.
However in $(2+1)$ dimensions the situation considerably simplifies, since the Wightman function of a scalar field conformally coupled to gravity in the background of a Banados-Teitelboim-Zanelli (BTZ) black hole can be given by a much simpler sum over images.

For this reason we specifically consider the BTZ black hole and investigate how Fisher information can be used to estimate the KMS temperature of the thermal response of a detector in its vicinity. We compare this to the corresponding case of a uniformly accelerating detector in its $(2+1)$-dimensional anti-de Sitter (AdS) counterpart, noting differences and similarities with the $(3+1)$-dimensional case.

In this paper, we find that $(2+1)$-dimensional AdS exhibits a range of Fisher information behaviour similar to those of $(3+1)$-dimensional AdS
\cite{DuMann2021}, though for a narrower range of parameters. Moreover, we also show that there are fewer distinct qualitative behaviours for both $(2+1)$- and $(3+1)$-dimensional AdS than was previously believed. As for our analysis of the Fisher information for the BTZ black hole, while we recover the same qualitative behaviours observed for AdS, they  are present for a much larger subset of parameter space. The discrepancy is sufficiently large to allow  for the effective discrimination between these two vacuum structures based solely on their Fisher information. In particular, the presence of local extrema in the Fisher information under a transparent boundary condition is not observed for AdS spacetimes, making it a sure indicator of a black hole in $(2+1)$ dimensions.

Our paper is structured as follows. First we will present some general formalism pertaining to Fisher information, the spacetimes of interest to us, and the UDW detector model. Then, in Section \ref{ch:udw} we  outline how we obtain Fisher information from our particular UDW set-up. Our results are then presented in Section \ref{ch:results} before our concluding remarks in the final section.

\section{General Formalism}\label{ch:formalism}

\subsection{Fisher Information}\label{subsec:FI}

Direct measurement of physical properties forms  the mainstay of classical physics. For many applications it is generally straightforward to measure quantities such as length, time, and temperature using appropriate classical measurement devices: rulers, clocks, or  thermometers. However at the frontier of physics, such measurements are often not practical, and more sophisticated measurement, or estimation, protocols are needed to extract the desired data. Fisher information is one of the more useful approaches to quantifying such protocols. It can be used as a measure when solving a parameter estimation problem in which we are asked to estimate the value of some non-observable parameter, $\xi$, based on the measured value of some observable parameter, $x$, where the relation between these two observables can be described by the probability distribution, $p(x|\xi)$.

Denote by $(X, \Xi)$   the set of possible values that $x$ and $\xi$ can respectively take. We define an estimator, $\hat{\xi}$, to be the function $\hat{\xi}: X^n \to \Xi$ that  returns an estimate for the underlying parameter $\xi$ given a sample of $n$ observables $x$. Such an estimator is said to be unbiased if the estimate returned is equal to the actual value of the underlying parameter.

Given an unbiased estimator, $\hat{\xi}$, the Fisher information, $\I(\xi)$, is 
\begin{equation}\label{eq:fisher-def}
	\I(\xi) = \int p(x|\xi) \left( \D{\ln p(x|\xi)}{\xi} \right)^2 dx = \int \frac{1}{p(x|\xi)} \left( \D{p(x|\xi)}{\xi} \right)^2 dx,
\end{equation} 
which quantifies the amount of information about the parameter $\xi$ carried by the observable $x$. 
The Cram\'er-Rao bound \cite{Rao1992,Cramer1999}
\begin{equation}\label{eq:crbound}
	\text{var}(\hat\xi) \geq \frac{1}{n\I(\xi)},
\end{equation}
then provides a lower bound on the variance of the estimator (equal to the mean-squared error for an unbiased estimator), specified by the Fisher information, $\I(\xi)$, and the number of measurements, $n$. Throughout this paper we assume that we are performing a single measurement and thus set $n=1$. The larger the Fisher information $\I(\xi)$, the lower the bound, and thus the more accurate an estimation we are able to make.

\subsection{Anti-de Sitter space and the BTZ black hole}\label{subsec:spacetimes}

The BTZ black hole spacetime can be obtained from identifications of $(2+1)$-dimensional AdS spacetime. AdS spacetime is the maximally symmetric spacetime with negative curvature and can be represented as an $(n+1)$-dimensional hyperboloid embedded in an $(n+2)$-dimensional flat spacetime.
 In our case, (2+1)-dimensional AdS with cosmological constant $\Lambda = -1/\ell^2$ can be expressed as the   hyperboloid 
\begin{equation}\label{eq:ads-hyp}
	X_1^2 + X_2^2 - T_1^2 - T_2^2 = - \ell^2,
\end{equation}
embedded in the  $(2+2)$-dimensional flat spacetime  
\begin{equation}\label{eq:minkowski22}
	dS^2=  dX_1^2 + dX_2^2 - dT_1^2 - dT_2^2.
\end{equation} 

Since we will be considering a constantly accelerating detector, we will make use of the AdS-Rindler metric 
\begin{equation}\label{eq:ads-metric}
	ds^2 = - \left( \frac{r^2}{\ell^2} - 1 \right) dt^2 + \left( \frac{r^2}{\ell^2} - 1 \right)^{-1}dr^2 + r^2d\phi^2,
\end{equation}
of constant negative curvature  obtained from \eqref{eq:ads-hyp} and \eqref{eq:minkowski22} via the transformations
\begin{eqnarray}\label{eq:ads-transformations}
	T_1 &=& \ell \sqrt{\frac{r^2}{\ell^2}} \cosh \phi \, ,
	\quad
	X_1 =   \ell \sqrt{\frac{r^2}{\ell^2}} \sinh \phi \, ,
	\nn\\
	T_2 &=& \ell \sqrt{\frac{r^2}{\ell^2}-1} \sinh \frac{t}{\ell} \, ,
	\quad
	X_2 = \ell \sqrt{\frac{r^2}{\ell^2}-1} \cosh \frac{t}{\ell} \, ,
\end{eqnarray}
for which we note the presence of an acceleration horizon at $r=\ell$ while $\phi$ takes on values from the real line.

We can then obtain the relevant formulae for the BTZ spacetime in an analogous fashion starting with the BTZ metric
\begin{equation}\label{eq:btz-metric}
	ds^2 = - \left( \frac{r^2}{\ell^2} - M \right) dt^2 + \left( \frac{r^2}{\ell^2} - M \right)^{-1}dr^2 + r^2d\phi^2,
\end{equation}
which is obtained from \eqref{eq:ads-hyp} and \eqref{eq:minkowski22} via the transformations
\begin{align}\label{eq:btz-transformations}
	T_1 &= \ell \sqrt{\frac{r^2}{M \ell^2}} \cosh (\sqrt{M} \phi) \, , \quad
	X_1 =   \ell \sqrt{\frac{r^2}{M \ell^2}} \sinh (\sqrt{M} \phi) \, ,
	\nn\\
	T_2 &= \ell \sqrt{\frac{r^2}{M \ell^2}-1} \sinh \frac{\sqrt{M} t}{\ell} \, ,    \quad
	X_2 = \ell \sqrt{\frac{r^2}{M \ell^2}-1} \cosh \frac{\sqrt{M} t}{\ell} \, .
\end{align}
followed by the identification $\phi \sim \phi + 2 \pi$.

The vacuum Wightman function for a (massless) conformally coupled scalar field in $(2+1)$-dimensional AdS is given by
\begin{equation}\label{eq:ads-wightman}
	W_\text{AdS}(x,x')=\frac{1}{4\pi\sqrt{2}\ell} \left(\frac{1}{ {\sqrt{\sigma(x,x')}}}-\frac{\zeta}{ {\sqrt{\sigma(x,x')+2}}}\right),
\end{equation}
where
\begin{align}
	\sigma(x,x') &= \frac{1}{2\ell^2} \left[ \left(X_1-X_1' \right)^2- \left(T_1-T_1'\right)^2 +\left(X_2-X_2'\right)^2 - \left(T_2-T_2'\right)^2 \right], 
\end{align}
is the square distance between $x$ and $x'$ in the embedding space $\mathbb{R}^{2,2}$.
The parameter $\zeta \in \{1,0,-1\}$ respectively specifies the  Dirichlet ($\zeta = 1$), transparent ($\zeta = 0$), and Neumann ($\zeta = -1$) boundary conditions satisfied by the field at spatial infinity \cite{AvisIshamStorey1978}.

The BTZ Wightman function for the Hartle-Hawking vacuum can be then constructed from the AdS Wightman function using the method of images \cite{LifschytzOrtiz1994}
\begin{align}
	W_{\rm BTZ}(x,x') &= \sum_{n=-\infty}^\infty  \, W_{\rm AdS}(x, \Gamma^n x'), 
\end{align}
where $W_{\rm AdS}(x, x')$ is the vacuum Wightman function associated with a massless conformally coupled scalar field in  AdS from  \eqref{eq:ads-wightman} and $\Gamma^n x'$ denotes the action of the identification  on the spacetime point $x'$.

This works out to be
\begin{align}\label{eq:btz-wightman}
	W_{\rm {BTZ}}(x,    x') =  \frac{1}{4 \pi \sqrt{2} \ell } \sum_{n=-\infty}^\infty\left[ \frac{1}{\sqrt{\sigma_n}} -  \frac{\zeta}{\sqrt{\sigma_n + 2}} \right],
\end{align}
where
\begin{align}
	\sigma_n(x,x') &\ce  \frac{r r'}{\rh^2} \cosh\! \left[\frac{\rh}{\ell} ( \Delta \phi - 2 \pi n) \right] -1 - \frac{\sqrt{ (r^2 - \rh^2)(r'^2 - \rh^2)}}{\rh^2}  \cosh \!\left[\frac{\rh}{\ell^2} \Delta t \right],   
\end{align}
in the coordinates \eqref{eq:btz-metric},  with $\Delta \phi \ce \phi-\phi'$ and $\Delta t \ce t-t'$.  The $n=0$ term is the Wightman function for AdS-Rindler space.



\subsection{Unruh-DeWitt detectors} \label{sec:udw}

We will make use of a two-level quantum particle detector known as an Unruh-DeWitt detector. The ground state, $\ket{0_D}$, and excited state, $\ket{1_D}$, are separated by an energy gap, $\Omega$. The detector moves along some trajectory, $x(\tau)$, parametrized by the proper time of the detector, $\tau$, and is coupled to a massless scalar field, $\phi$ as described by the interaction Hamiltonian
\begin{equation}\label{eq:udw-ham}
	H_I = \lambda  \chi(\tau) \left( e^{i\Omega\tau} \sigma^+ + e^{-i\Omega\tau} \sigma^- \right) \otimes \phi[x(\tau)],
\end{equation}
where $\lambda$ is the coupling constant, and $\sigma^+ = \ket{1_D}\bra{0_D}$ and $\sigma^- = \ket{0_D}\bra{1_D}$ are ladder operators on the Hilbert space of the detector. The switching function $\chi(\tau)$ quantifies both the strength and duration of  time that the detector couples to the field.
We shall set $\chi(\tau) = 1$ henceforth, as we are working in the context of an open quantum system where the detector always couples to the field.

In a perturbative regime, to the lowest order in $\lambda$, the transition probability from $\ket{0_D}$ to $\ket{1_D}$ is proportional to the response function  
\begin{equation}
\label{F-response}
	F(\Omega) = \int_{-\infty}^\infty d\tau \int_{-\infty}^\infty d\tau'\, e^{-i\Omega(\tau-\tau')}\, W(x(\tau),x(\tau')),
\end{equation}
where $W(x(\tau),x(\tau'))$ is the Wightman function pertinent to the spacetime under consideration. In general it quantifies the correlations in the field between the points $x(\tau)$ and $x(\tau')$ dependent on the trajectory of the detector. For a stationary trajectory 
 the Wightman function only depends on $\Delta \tau = \tau - \tau^\prime$. For such trajectories, we can consider the response per unit time \cite{Jennings2010,LoukoSatz2008}, which can be thought of as the instantaneous change in the response function, and is given by
\begin{equation}\label{eq:response-rate}
	\F(\Omega) = \int_{-\infty}^\infty d\Delta \tau \; e^{-i\Omega\Delta\tau}\, W(\Delta\tau).
\end{equation}
This response rate plays a special role when considering UDW detectors as open quantum systems.

Our interest is in the UDW detector's ability to act as a thermometer of the spacetime (measuring thermality), and especially their ability to discriminate between different quantum vacuum structures \cite{Langlois:2004fv,Langlois:2005nf,Smith:2013zqa,Martin-Martinez:2015qwa,Gray:2021dfk,Henderson:2022oyd}. 
We observe thermal states in expanding dS spacetime \cite{BirrellDavies1984, GibbonsHawking1977}, for uniform detector acceleration, or more generally whenever there is an event horizon \cite{Sewell1982}. It is well known that a UDW detector near a black hole will experience thermal radiation \cite{Hawking1975}, but it is also the case that such a detector will experience thermal radiation in AdS provided that it has a sufficiently large acceleration \cite{Jennings2010,DeserLevin1997}.

We will consider how Fisher information can be used to estimate the thermal response of a stationary UDW detector outside of a BTZ black hole.
We begin by first considering a simpler version of this problem, namely an accelerated detector in $(2+1)$-dimensional AdS. In both cases the response rate \eqref{eq:response-rate} has a known analytical result.

\subsubsection{Accelerated detector in AdS}

Inserting the AdS Wightman function from  \eqref{eq:ads-wightman} within the response rate formula found in   \eqref{eq:response-rate}, we find that the response rate of a uniformly accelerating detector in $(2+1)$-dimensional AdS is    
\begin{align}
	\F_{\rm AdS} =  \frac{1}{4} - \frac{i}{4 \pi} {\rm PV} \int_{-\infty}^\infty dz \frac{e^{- i \Omega  z/ (\pi T)}}{\sinh z} - \frac{\zeta}{2 \pi\sqrt{2}} {\rm Re}  \int_{0}^\infty dz \frac{e^{-i \Omega z/ (2 \pi T)}}{\sqrt{1 + 8 \pi^2 \ell^2 T^2 - \cosh z }} ,
	\label{RindResponse}
\end{align}
which, after performing the integrals, can be written as 
\begin{align}\label{eq:ads-responserate}
	\F_{\rm AdS} = \frac{1}{4} \left[1 - \tanh \left(\frac{\Omega}{2 T} \right) \right] \times\left\{1   - \zeta  P_{-\frac{1}{2} + \frac{i \Omega}{2 \pi T}} \left(1 + 8 \pi^2 \ell^2 T^2 \right)  \right\} ,
\end{align}
where $P_{\nu}$ is the associated Legendre function of the first kind, satisfying $P_{-1/2 + i \lambda} = P_{-1/2 - i \lambda}$, and $T$ is the KMS temperature of the vacuum. We note that the response rate 
\eqref{eq:ads-responserate} 
is curiously the same as that of the response function \eqref{F-response}
with a Gaussian switching function of infinite width  \cite{HHMSZ2020} apart from a factor of $\sqrt{\pi}$.

For a uniformly accelerated detector in AdS space, this temperature can be expressed as a function of the detector's acceleration, $a$, and the AdS length, $\ell$ \cite{HHMSZ2020}:
\begin{equation}\label{eq:ads-temperature}
	T = \frac{\sqrt{a^2\ell^2-1}}{2\pi\ell}.
\end{equation}

It is worth noting that while an accelerated detector in Minkoswki spacetime experiences a temperature $T=a/2\pi$ for all accelerations $a>0$, in AdS spacetime there is a thermal response only when $a>1/\ell$, i.e., the acceleration is larger than the inverse AdS length. This condition is ensured 
in the AdS-Rindler coordinate system \eqref{eq:ads-metric}.

 \subsubsection{Stationary detector in BTZ}

If we instead insert the BTZ Wightman function from   \eqref{eq:btz-wightman} into the response rate formula, we obtain
\begin{align}
	\F_{\rm BTZ} = \F_{\rm AdS} +&\frac{1}{\sqrt{2 \pi}} \sum_{n=1}^{\infty} \Bigg\{\int_{0}^{\infty} dz\ {\rm{Re}}\Bigg[ \frac{\exp\big(-i \Omega z/(2 \pi T) \big)}{\sqrt{\cosh\big(\alpha_{n}^-\big)-\cosh(z)}}\Bigg] 
	\nn\\
	&- \zeta\int_{0}^{\infty} dz\ {\rm{Re}}\Bigg[\frac{\exp\big(-i\Omega z/(2 \pi T)\big)}{\sqrt{\cosh\big(\alpha_{n}^+\big)-\cosh(z)}}\Bigg]\Bigg\} \, ,	\end{align}
for a stationary detector in the  BTZ spacetime, with
	\begin{align}
		\cosh \alpha^\mp_n &=  \left(1 + 4 \pi^2 \ell^2 T^2 \right)\cosh (2 \pi n \sqrt{M})  \mp 4 \pi^2 \ell^2 T^2 \, .
	\end{align}
and where the AdS-Rindler result $\F_{\rm AdS}$ \eqref{eq:ads-responserate} arises as the $n=0$ term of \eqref{eq:btz-wightman} while the terms from $n=-\infty,...,-1$ can be expressed alongside their positive counterparts by symmetry. 
Hence the remaining terms ranging from $n=1,...,\infty$ constitute the novel black hole effects. Note that these additional terms are parametrized by the black hole mass, $M$, with more pronounced effects arising for smaller $M$.

Computing the integrals, we can rewrite the response rate as
\begin{align} 
	\F_{\rm BTZ} &= \frac{1}{4} \left[1 - \tanh\left(\frac{\Omega}{2 T}\right) \right] \sum_{n=-\infty}^{n = \infty} \left[P_{-\frac{1}{2} + \frac{i \Omega}{2 \pi T}} \left( \cosh \alpha_n^- \right) - \zeta  P_{-\frac{1}{2} + \frac{i \Omega}{2 \pi T}} \left( \cosh \alpha_n^+ \right) \right].
\end{align}
For a stationary detector outside the BTZ black hole, the temperature is also given by \eqref{eq:ads-temperature} as in AdS \cite{HHMSZ2020}.

\section{Metrology with Unruh-DeWitt detectors} \label{ch:udw}

Throughout this paper, we will make use of an open quantum systems framework applied to  UDW detectors \cite{DuMann2021,HFZF2018, BenattiFloreanini2004,TWFJ2015}. This will allow us to consider the evolution of the system coupled to the vacuum before tracing out the field to examine the state of the subsystem of interest, the UDW detector. With the appropriate assumptions, the evolution of our system can in fact be isolated.

To begin, we define the overall Hamiltonian of the combined system, including detector and the quantum field, in addition to the interaction Hamiltonian, to be
\begin{equation}
    H = H_D + H_\phi + H_I,
\end{equation}
where $H_D = \frac12 \Omega a_D^\dagger a_D = \frac{1}{2}\Omega(\ket{0_D}\bra{0_D} - \ket{1_D}\bra{1_D})$ is the free Hamiltonian of the detector with energy gap $\Omega$, $H_\phi= \frac{dt}{d\tau}\sum_\mathbf{k} \omega_\mathbf{k} a_\mathbf{k}^\dagger a_\mathbf{k}$ is the free Hamiltonian of the massless scalar field $\phi(x)$,  where the redshift factor $\frac{dt}{d\tau}$ allows us to time evolve with respect to the detector's proper time \cite{EduRickBruno2020,EduPablo2018}, and $H_I$ is the Unruh-DeWitt interaction Hamiltonian defined in \eqref{eq:udw-ham}.

The von Neumann equation dictates time evolution of the combined system as
\begin{equation}
    \D{\rho_\text{tot}}{\tau} = -i [H, \rho_\text{tot}],
\end{equation}
where $\rho_\text{tot}$ is the density matrix of the combined system, initialized in the state $\rho_\text{tot}(0) = \rho_D(0) \otimes \ket{0_\phi}\bra{0_\phi}$, where $\rho_D(0)$ is the initial state of the detector, and $\ket{0_\phi}$ is the conformal vacuum of the scalar field $\phi(x)$.
 One can obtain the state of the detector by taking the partial trace over the field of the combined state, i.e., $\rho_D = \tr_\phi \rho_\text{tot}$. 
 
If we assume a weak coupling ($\lambda \ll 1$) with field correlations decaying sufficiently fast for large time separations, then the density operator of the detector's evolution is expressed by the master equation of Kossakowski-Lindblad form 
\cite{BenattiFloreanini2004}. This is the most general description of Markovian time evolution of a quantum system \cite{Manzano2020}, and is given by
\begin{equation}\label{eq:lindblad}
    \D{\rho_D(\tau)}{\tau} = -i[H_\text{eff}, \rho_D(\tau)] + L[\rho_D(\tau)],
\end{equation}
where $H_\text{eff} = \frac12\Tilde{\Omega}(\ket{0_D}\bra{0_D} - \ket{1_D}\bra{1_D})$ is the effective Hamiltonian, and
\begin{equation}
    L[\rho] = \frac12 \sum_{i,j=1}^3 C_{ij} \left( 2\sigma_j\rho\sigma_i - \sigma_i\sigma_j\rho - \rho\sigma_i\sigma_j\right),
\end{equation}
where the $\sigma_i$ are the Pauli matrices. The quantity $\Tilde{\Omega}$ is a renormalized gap given by 
\begin{equation}
    \Tilde{\Omega} = \Omega + i \left[ \mathcal{K}(-\Omega) - \mathcal{K}(\Omega) \right],
\end{equation}
where $\mathcal{K}(\Omega)$ is the Hilbert transform of the response per unit time $\F(\omega)$ defined by
\begin{equation}
    \mathcal{K}(\Omega) = \frac{1}{i\pi} \text{ PV} \int_{-\infty}^\infty d\omega \, \frac{\F(\omega)}{\omega - \Omega},
\end{equation}
with PV denoting the Cauchy principal value. $C_{ij}$ is called the Kossakowski matrix, and is also completely determined by the response per unit time $\F(\Omega)$:
\begin{equation}
    C_{ij} = \begin{pmatrix}
    A & -iB & 0 \\
    iB & A & 0 \\
    0 & 0 & A+C
    \end{pmatrix},
\end{equation}
where
\begin{equation}\label{eq:A}
    A = \frac12[\F(\Omega) + \F(-\Omega)]
\end{equation}
\begin{equation}\label{eq:B}
    B = \frac12[\F(\Omega) - \F(-\Omega)]
\end{equation}
\begin{equation}
    C = \F(0) - A
\end{equation}

We note that \eqref{eq:lindblad} can be solved analytically. 
Given a detector initialized in the general pure state $\ket{\psi_D} =
\cos\frac{\theta}{2} \ket{0_D} + \sin\frac{\theta}{2} \ket{1_D}$, its density matrix at time $\tau$ is specified by the Bloch vector $\vec{a} = (a_1,a_2,a_3)$ such that
\begin{equation}\label{eq:det-state}
    \rho(\tau) = \frac12 \left( I + \vec{a}(\tau) \cdot \vec{\sigma}  \right),
\end{equation}
where $\vec{\sigma} = (\sigma_1,\sigma_2,\sigma_3)$ are the Pauli matrices, and the Bloch vector components are given by
\begin{equation}
    a_1(\tau) = e^{-A\tau/2} \sin\theta\cos\Tilde{\Omega}\tau,
\end{equation}
\begin{equation}
    a_2(\tau) = e^{-A\tau/2}\sin\theta\sin\Tilde{\Omega}\tau,
\end{equation}
\begin{equation}\label{eq:az}
    a_3(\tau) = -e^{-A\tau} \cos\theta - R(1-e^{-A\tau}).
\end{equation}
where $R = B/A$. Note that $|\vec{a}| < 1$ in general, implying that the evolution is non-unitary.

Having identified our state of interest and its time evolution dynamics, we are now in a position to compute the Fisher information for estimating temperature in AdS-Rindler space and the BTZ black hole with UDW detectors. Our estimation strategy is to first let the detector interact with a massless scalar field in the spacetime of interest then make a projective measurement of the detector's state after some detector proper time $\tau$.

Since we are working with a two-level quantum detector, the UDW detector has associated to it a 2-dimensional Hilbert space. It then follows that every measurement has two possible outcomes. By definition of the Fisher information \eqref{eq:fisher-def}, our generally continuous probability distribution $p(x|\lambda)$ reduces to the discrete probability of getting either outcome, and the integral simplifies to a two term sum.
 
If said outcomes occur with probabilities $p$ and $1-p$ then the Fisher information can be expressed as
\begin{equation}
    \I(\lambda) = \frac{1}{p} \left( \D{p}{\xi} \right)^2 + \frac{1}{1-p} \left( -\D{p}{\xi} \right)^2 = \frac{1}{p(1-p)} \left( \D{p}{\xi} \right)^2.
\end{equation}

When measuring the detector in a state specified by the Bloch vector $\vec{a}$ using the computational basis, $\{ \ket{0_D}, \ket{1_D} \}$, the probability of obtaining $\ket{0_D}$ is 
\begin{equation}
    p = \tr(\rho\ket{0_D}\bra{0_D})  = \frac12(1+a_3),
\end{equation}
and the probability of obtaining $\ket{1_D}$ is $1-p = \frac12(1-a_3)$. 

The Fisher information is thus given by
\begin{equation}\label{eq:fisher}
    \I(\xi) = \frac{(\partial_\xi a_3)^2}{1-a_3^2},
\end{equation}
where $a_3$ is the third component of the Bloch vector of the detector's state, given by \eqref{eq:az}. 

Since the parameter of interest for us is temperature, we set $\xi$ to be $T$, leading to the Fisher information
\begin{equation}\label{eq:fisher-t}
    \I(T) = \frac{(\partial_T a_3)^2}{1-a_3^2}.
\end{equation}
We shall find it convenient to rescale $\I(T)$ by $T^2$ in order to more easily compare for various $T$ . We note that this $\I=\I(T)T^2)$, which will simply refer to as the Fisher information is unitless.

We will now identify all of the parameters that will be featured in our upcoming analysis, beginning with the AdS length, $\ell$, which we use as the base units for all other parameters. 
The first two parameters of note are the temperature experienced by the detector in the vacuum, $T$, and the energy gap of the detector $\Omega$, which generally appear in the dimensionless ratio  $\Omega/T$, though when appearing separately they will appear as $T\ell$ and $\Omega\ell$.  

The remaining relevant parameters are the initial state of the detector, $\theta$, which we will often fix to one of $\{0, \pi/2, \pi\}$, 
the detector interaction time $\tau$, which is $\tau/\ell$ in its unitless form, and the spacetime boundary condition, $\zeta \in \{0, 1, -1\}$ corresponding to transparent, Dirichlet, and Neumann boundary conditions respectively. Finally, in the BTZ spacetime, the remaining parameter is the dimensionless mass $M$, which is unitless.

\section{Results}\label{ch:results}

In this section, we will first discuss the Fisher information of $(2+1)$-dimensional  AdS-Rindler, and compare it with  previous results for the $(3+1)$-dimensional AdS and dS cases \cite{DuMann2021}.   We shall then discuss the results for the BTZ spacetime, which is an extension of the AdS-Rindler case via the image sum.  

Our goal is to  identify qualitative behaviours of the Fisher information for a given spacetime. This motivation is two-fold.
First, this will allow us to identify better thermal estimation procedures. Given a specific spacetime and some known detector parameters (energy gap, initial state) we can identify the time (early, middle, or late) at which we can obtain the best estimate of the temperature.  
Second, we also want be able to use the Fisher information as a probe of the underlying spacetime. By this we mean to ask, given a particular Fisher information behaviour, can we discriminate between the possible background spacetimes?

\subsection{AdS-Rindler  Spacetime}
 
We begin with the case of a uniformly accelerating detector in $(2+1)$-dimensional AdS.  As previously noted, to experience thermal excitation the detector's acceleration  must be supercritical, with $a \geq 1/\ell \equiv k$.  Accelerations beneath this supercritical value have zero response rate.
 
\begin{figure}
	\centering
	\includegraphics[width=\linewidth]{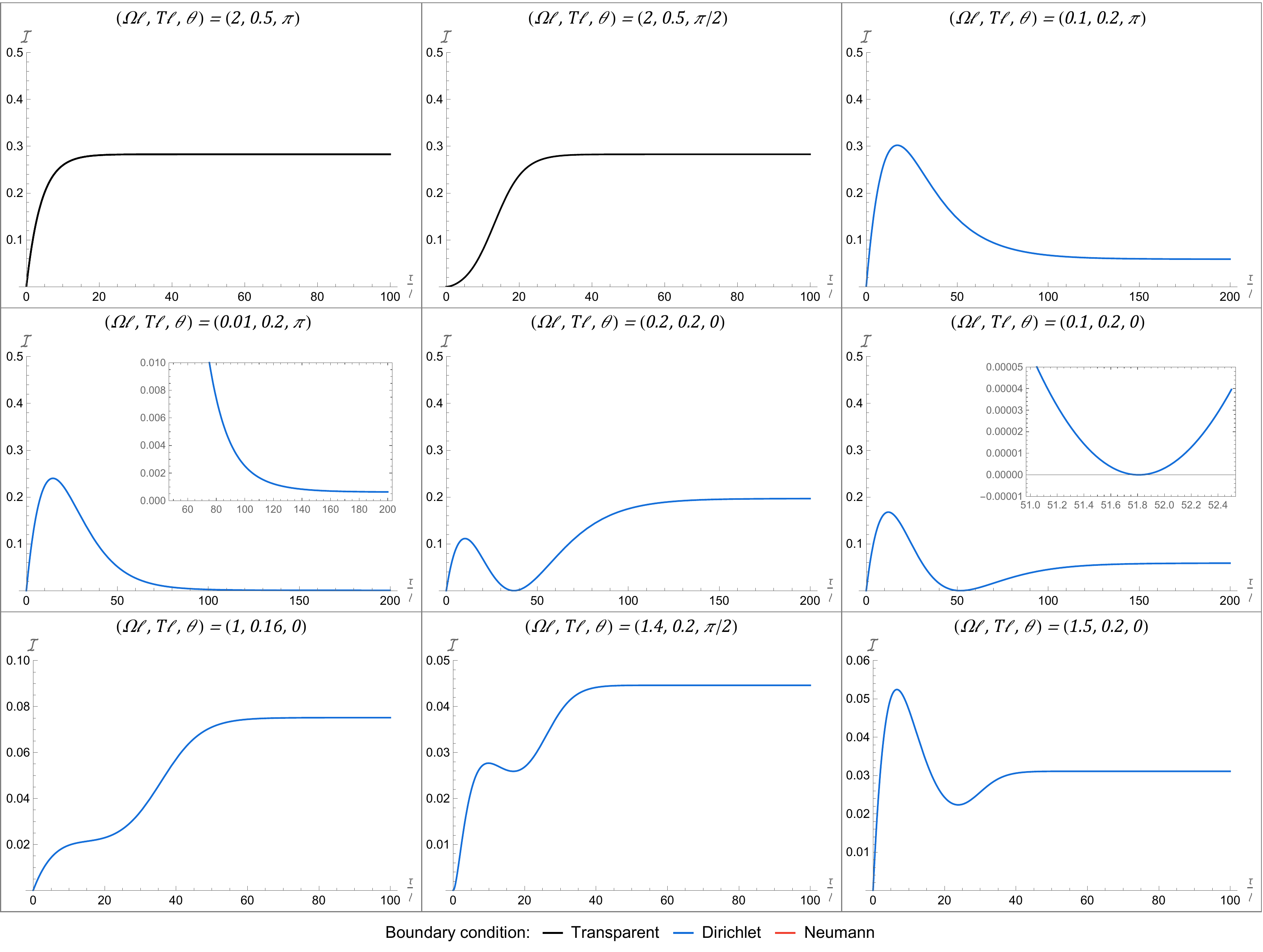}
	\caption[Short caption - doesn't seem to appear anywhere]{This grid plot of the temporal Fisher information displays all the distinct behaviours observed for the AdS Rindler case. 
	From left-to-right and top-to-bottom, we label these as behaviours 1 through 9. There are eight distinct behaviours, with behaviours 3 and 4 being qualitatively the same.
    We note that while both the Dirichlet and Neumann boundary conditions ($\zeta=1$ and $\zeta=-1$ respectively) can be used to obtain all nine behaviours, only the first two can be obtained from the more fundamental transparent boundary condition ($\zeta=0$). To highlight this fact, we have plotted the first two behaviours using the transparent boundary condition (in black), while the remaining six qualitatively distinct behaviours were plotted using the Dirichlet boundary condition (in blue).
	}
	\label{fig:ads-qb-grid}
\end{figure}

Plotting the Fisher information as a function of time in Figure \ref{fig:ads-qb-grid}, we observe nine distinct qualitative behaviours.  These are analogous to those seen for a uniformly accelerating detector in  $(3+1)$-dimensional AdS \cite{DuMann2021}. However there are a few noteworthy observations to be made.

Numbering the subfigures in Figure \ref{fig:ads-qb-grid} as on a telephone keypad, we see that behaviours   3 and 4   are qualitatively the same. 
The inset plot for  behaviour 4 does not asymptote to zero but rather to a small but finite positive value, qualitatively the same as behaviour 3. This double-counting of behaviours is also present in \cite{DuMann2021}, leaving only eight truly distinct behaviours in AdS-Rindler in both $(2+1)$ and $(3+1)$  dimensions.   We have also verified that  the minima in behaviours 5 and 6  do indeed attain 0, at least to high numerical certainty, as shown in the inset plot of behaviour 6.

Although we recover the same eight distinct qualitative behaviours in $(2+1)$-dimensions as were found in $(3+1)$-dimensions, only two of these are attainable using the transparent, boundary condition, whereas   five distinct behaviours were attainable from the transparent boundary condition in $(3+1)$-dimensional AdS-Rindler.  Hence there is a broader range of freedom in Fisher information behaviour in the $(3+1)$-dimensional AdS-Rindler spacetime, corresponding to a bosonic response rate distribution, compared to its $(2+1)$-dimensional counterpart, which corresponds to a fermionic response rate distribution. In both instances, the remaining behaviours were both attainable via the Dirichlet and Neumann boundary conditions, though only the Dirichlet boundary conditions are displayed in Figure \ref{fig:ads-qb-grid}.

If we turn our attention to the overall trends in the  temporal behaviour of the Fisher information, we note that it always begins by growing from zero, and it always ends by plateauing to an asymptotic limit. This agrees with the intuition that before our detector interacts with the field, it has no information about the field, and that after interacting with the field for a sufficiently large time, it will have thermalized and thus has no more information to extract about the temperature. 

What is most intriguing is the behaviour between these start and end points. There appears to be anywhere from 0 to 2 inflection points, which characterize the intermediate-time behaviour. In some cases, the Fisher information is monotonically increasing as seen in behaviours 1, 2, and even 7 where there are two inflection points. The remaining behaviours all have a local maximum (which might also be a global maximum), while some temporal Fisher information behaviours  also have a local minimum.
It is worth noting that  the optimal thermal estimation procedure should be implemented before thermalization at the time corresponding to the global maximum, should one be present.  

Alternatively, it is possible for the local minimum to reach a value of zero at some intermediate  time. This effectively means that it is impossible to estimate the temperature at that time, despite there being non-zero Fisher information both before and after this minimum. It is unclear what physical intuition might be attributed to this behaviour. 

\begin{figure}[h]
	\centering
	\includegraphics[width=\textwidth]{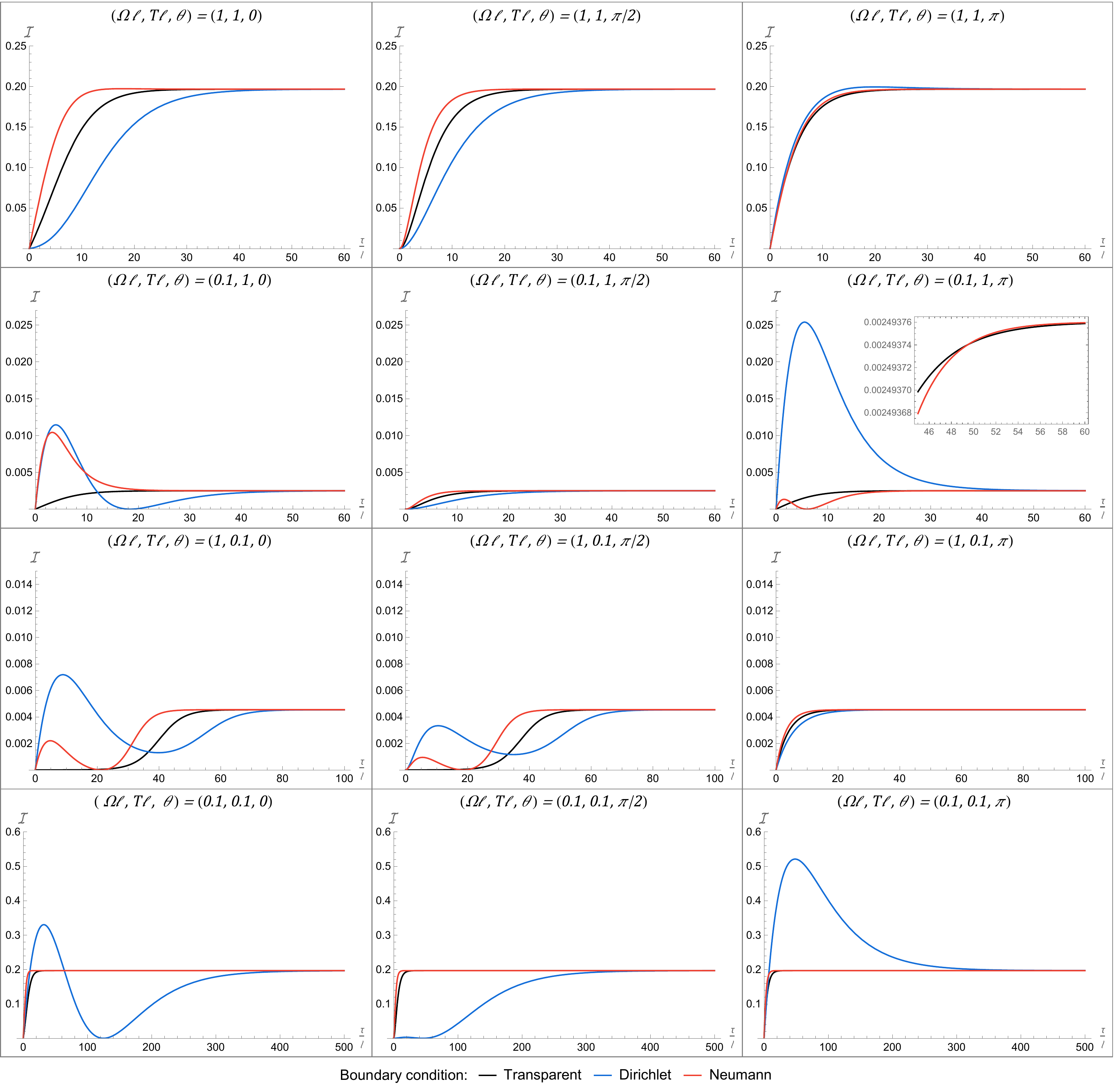}
	\caption[Short caption]{This grid plot displays the temporal Fisher information in AdS for all boundary conditions for various parameter choices. Each row of this grid plot fixes a pair of temperature and energy gap, $(T,\Omega)$, while each column fixes an intial state of the detector, $\theta$. Across all plots the colour of the curves represent the boundary condition.}
	\label{fig:ads-bc-grid}
\end{figure}

  The perspicacious reader might observe from Figure \ref{fig:ads-qb-grid} that the asymptotic value of the Fisher information seems to not be dependent on the initial state of the detector. This observation will be made more apparent in   Figure \ref{fig:ads-bc-grid}, where we compare the different boundary conditions in  a subset of the phase space.  Here, we observe a number of interesting behaviours.

First, it is immediately apparent that regardless of the boundary condition or initial state, there is a common asymptotic value for the Fisher information at late times given a particular set of parameters. This value can easily be computed analytically and is very much dependent on the Kossakowski ratio, $R$. 
Noting that $\lim_{\tau \to \infty} a_3 = -R$, we obtain
\begin{equation}
	\lim_{\tau \to \infty} \I(T) = \lim_{\tau \to \infty} \frac{(\partial_T a_3)^2}{1-a_3^2} = \frac{(\partial_T R)^2}{1-R^2},
\end{equation}
and so  compute 
\begin{equation}\label{KossRatio}
	R^{\text{AdS}} = -\tanh\left( \frac{\Omega}{2T} \right)
\end{equation}
for the value of the Kossakowski ratio, 
making use of the fact that the hyperbolic tangent found in   \eqref{eq:ads-responserate} is odd, and the fact that the Legendre polynomial satisfies $P_{-\frac12+i\lambda} = P_{-\frac12-i\lambda}$.  Surprisingly, this expression for the Kossakowski ratio is identical to that in $(3+1)$ dimensions. 

The asymptotic Fisher information is then
\begin{equation}\label{eq:asym}
	\I_{\text{asym}}^{\text{AdS}} = \frac{\Omega^2}{4T^2} \,\sech^2\left(\frac{\Omega}{2T}\right)
\end{equation}
which depends only the ratio $\Omega/T$.  Both the first and last row of Figure \ref{fig:ads-bc-grid} illustrate this property: 
despite these rows having different values parameterizing the temperature and energy gap, the ratio $\Omega/T$ is the same for both.
The expression \eqref{eq:asym} is the same in $(2+1)$ and $(3+1)$ dimensions, and so the value of  $\Omega/T$ ratio that maximizes 
$\F_{\text{asym}}^{\text{AdS}}$  is $\tanh\left( \frac{\Omega}{2T} \right) = \frac{2T}{\Omega}$  \cite{DuMann2021}. We will also see in our next subsection that these trends extend to the BTZ spacetime.

Second, we find that Fisher information with transparent boundary conditions is always approached asymptotically from above by its counterpart for Neumann boundary conditions and from below by its counterpart for  Dirichlet boundary conditions. In particular, the inset plot in Figure \ref{fig:ads-bc-grid}   shows the Neumann boundary condition overtaking the transparent one before they both asymptote to a common value. 

Summarizing, we find that  the overall structure of Fisher information in $(2+1)$-dimensional AdS-Rindler, though narrower in scope, is for the most part akin to  its $(3+1)$ dimensional counterpart.  The qualitative behaviours are similar as is the asymptotic behaviour.

\subsection{BTZ Spacetime}

We now turn to consideration of the Fisher information for the BTZ black hole. By  comparing these results to the AdS-Rindler case in the previous subsection we can identify which behaviours can truly be attributed to the black hole nature of this spacetime.

\begin{figure}[h!]
	\centering
	\includegraphics[width=\linewidth]{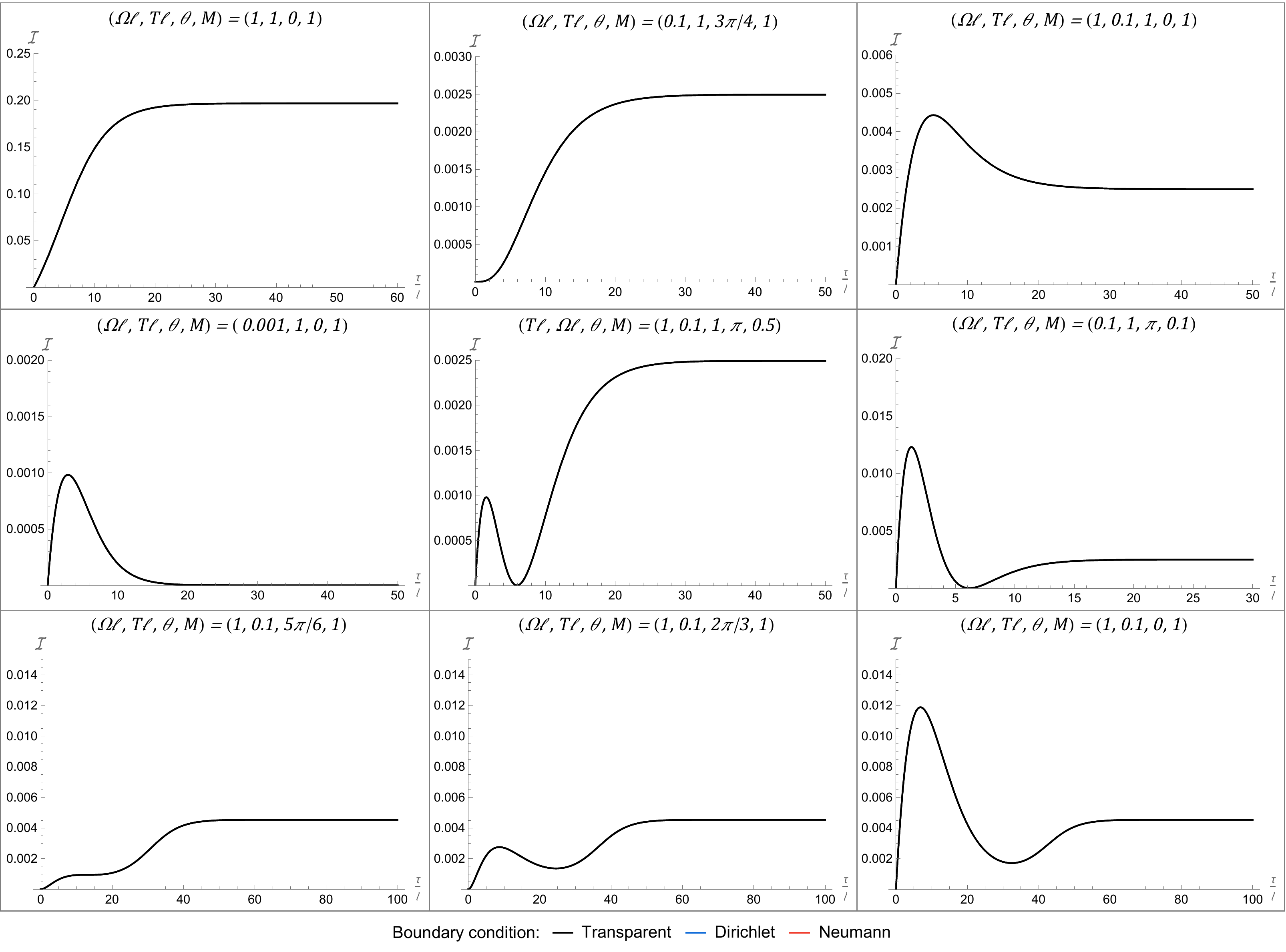}
	\caption[Short Caption]{All previously identified qualitative behaviours can be reproduced in the BTZ spacetime.  All of the behaviours displayed here were obtained using only the transparent boundary condition (in black). Each can also be   achieved using either Dirichlet or Neumann boundary conditions.}
	\label{fig:btz-qb-grid}
\end{figure}

In Figure \ref{fig:btz-qb-grid}, we plot the Fisher information as a function of time for the BTZ black hole. We find the same eight qualitatively distinct behaviours previously identified for AdS-Rindler in Figure \ref{fig:ads-qb-grid}. Nevertheless there are a few noteworthy points to be made.

First, unlike the AdS-Rindler case, all eight behaviours are present using only   transparent boundary conditions (as well as for Neumann and Dirichlet boundary conditions, as one might expect). This seems to hint at some additional complexity that is absent from the AdS spacetime.  In particular, if we observe any of the behaviours 3 through 8 with a transparent boundary condition, we know that our detector must be stationary in a BTZ spacetime and not accelerating in AdS.  We have set $M=1$ for all subfigures, with the exception of behaviours 5 and 6, where $M=0.5$ and $M=0.1$ better illustrate these particular behaviour, though they are also present for $M=1$.
 
\begin{figure}[h]
	\centering
	\includegraphics[width=\linewidth]{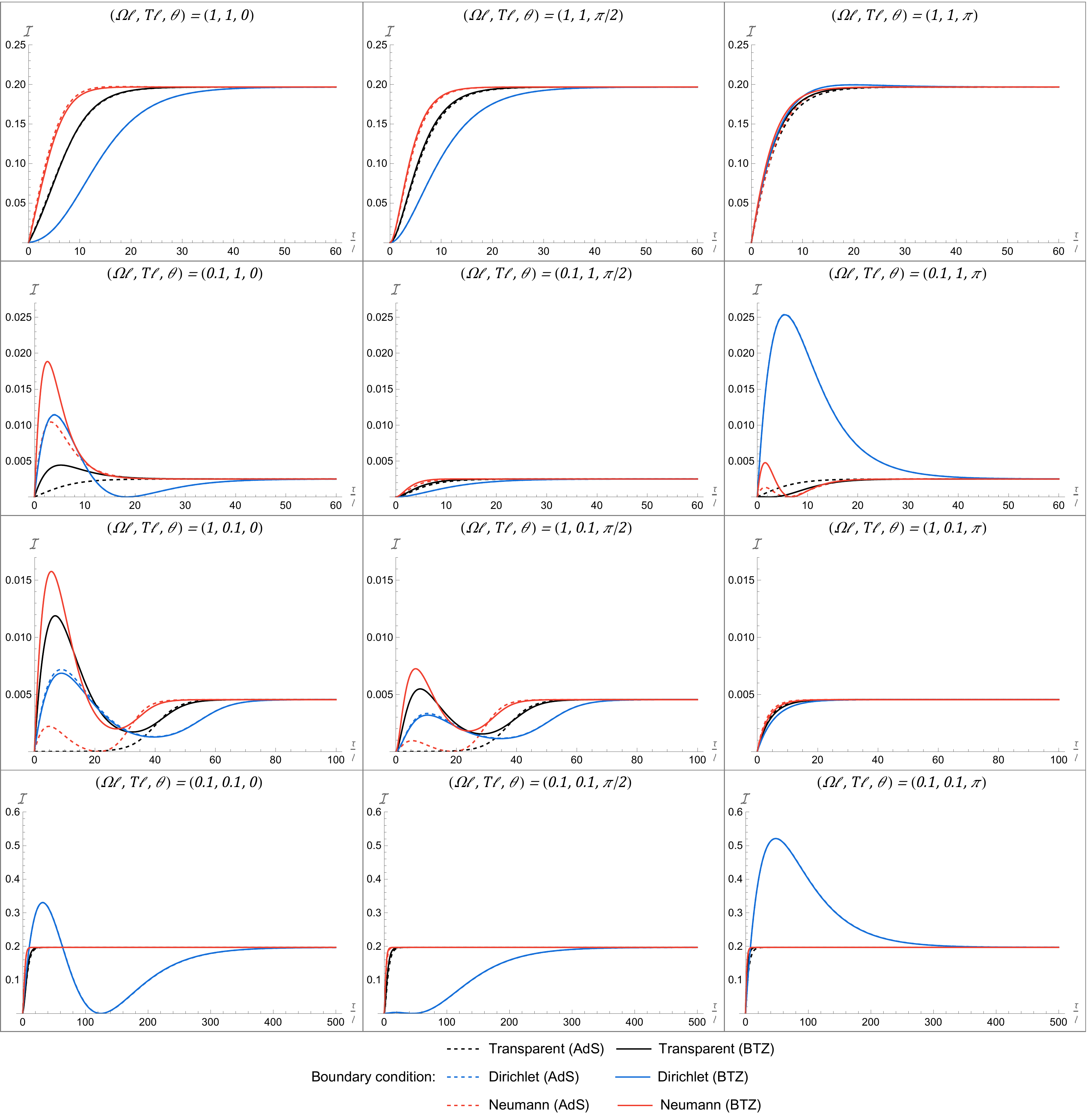}
	\caption{Here we directly compare the Fisher information in BTZ (solid lines) with the Fisher information in AdS (dashed lines) that was previously presented in Figure \ref{fig:ads-bc-grid}. The layout and choice of parameters is the same: the initial state of the detector, $\theta$, varies across columns, while the temperature, $T\ell$, and energy gap, $\Omega\ell$,  vary from row to row. The mass parameter for the BTZ spacetime is set to $M=1$ throughout this grid. }
	\label{fig:ads-vs-btz-bc-grid}
\end{figure}

We compare in Figure \ref{fig:ads-vs-btz-bc-grid} the Fisher information in the AdS-Rindler and  BTZ spacetimes, taking the same subset of parameter space as previously considered in Figure \ref{fig:ads-bc-grid}. We set the BTZ mass parameter to be $M=1$ since that allows the transformations  \eqref{eq:btz-transformations} to be identical to those in  \eqref{eq:ads-transformations} modulo the identification in $\phi$. 

We see that for the first and final rows, in which $\Omega=T$, the AdS and BTZ behaviours are essentially identical. However in rows two and three, which we can think of as the `hot' and `cold' environments respectively, we can distinguish between the two spacetimes. In the `hot' case, with $T\ell=1$ and $\Omega\ell=0.1$, the AdS-Rindler and BTZ curves overlap for Dirichlet boundary conditions, whereas a modest change appears for transparent boundary conditions. The Neumann boundary condition exhibits an amplification of its maximum in the BTZ case. The most dramatic changes are observed in the `cold' case, with $T\ell=0.1$ and $\Omega\ell=1$. We observe significant amplification (or even appearance) of maxima in the BTZ case at early times for all boundary conditions, except  at $\theta=0$, where the BTZ case has a slightly smaller maximum for Dirichlet boundary conditions. In particular, the conjunction of a  maximum in the Fisher information with the transparent boundary condition allows us to distinguish between the BTZ and AdS-Rindler spacetimes, since there is no such maximum present in AdS-Rindler with the transparent boundary condition. The BTZ curves also develop minima at intermediate times before asymptotically approaching (from below) their final values, which are the same as the AdS-Rindler case.  As such, after the detector has thermalized at late time, it is no longer able to distinguish between AdS-Rindler and BTZ spacetimes.
  
We next consider what influence the BTZ mass parameter $M$ has on the Fisher information. We find that for some choices of detector parameters, the Fisher information changes in an erratic fashion as $M$ varies, whereas  for others it varies monotonically.

\begin{figure}
	\centering
	\includegraphics[width=\linewidth]{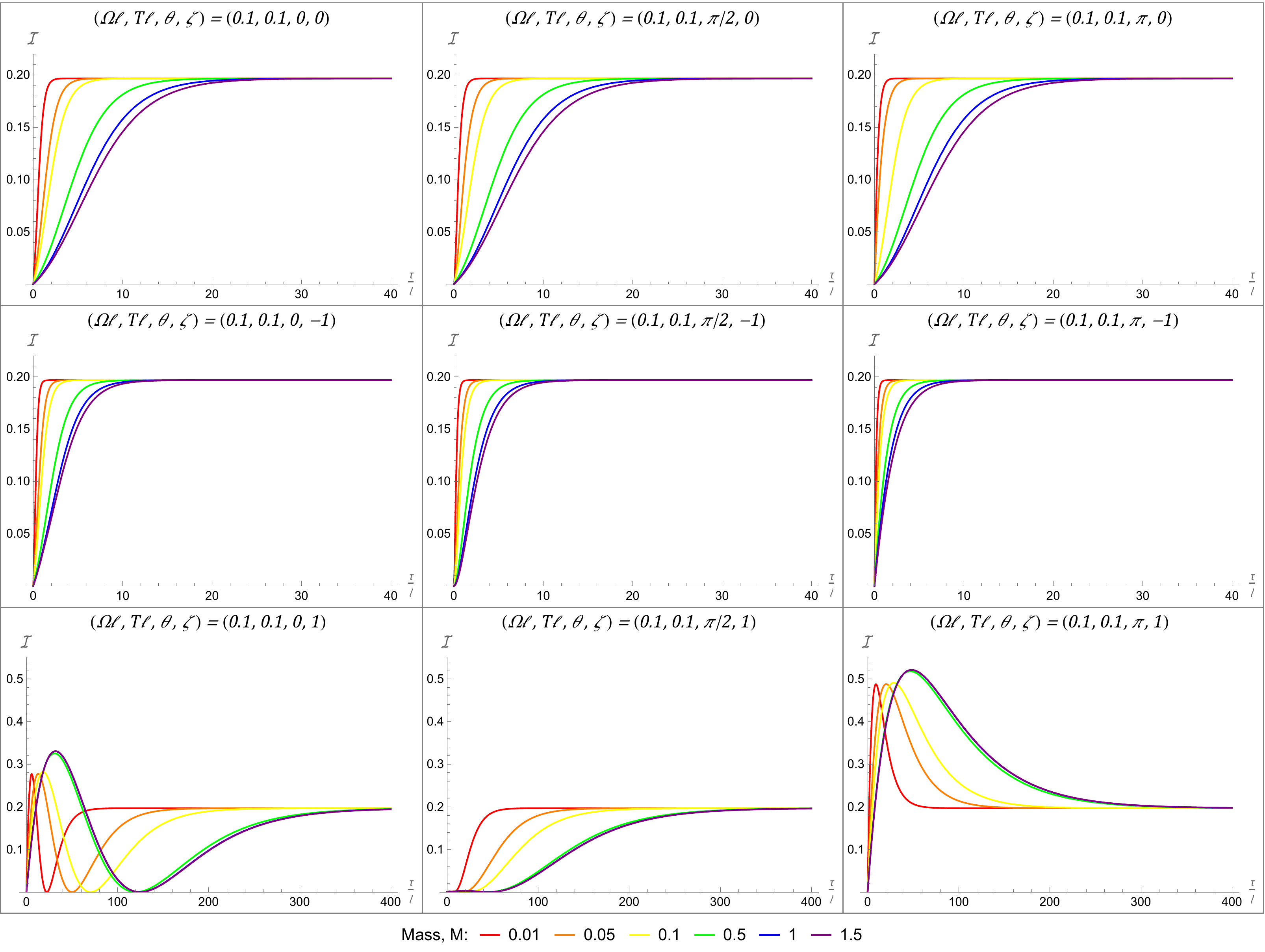}
	\caption[short caption]{This grid of plots allows us to identify how the Fisher information changes under variations of the mass, $M$. The mass is increasing along the colours of the rainbow, from $M=0.01$ to $M=1.5$. The energy gap and temperature are fixed throughout the grid. Each column represents a different initial state, while each row represents a different boundary condition. We note that for $\Omega = T = 0.1$, a variation of the mass parameter does not seem to have a substantial effect on the Fisher information. Increasing mass seems to delay the increase in Fisher information, with a slight increase in the maximum value for the Dirichlet boundary condition.}
	\label{fig:btz-varyingm-from0101}
\end{figure}

As expected from Figure \ref{fig:ads-vs-btz-bc-grid}, when $\Omega=T$ a small change in $M$ yields a small change in the Fisher information. 
We show a comparison for various values of $M$ in Figure \ref{fig:btz-varyingm-from0101} where, similar to previous grids, each column depicts a given initial state, and we have encoded the boundary conditions into the rows of the grid. 
As we decrease $M$ by over two orders of magnitude, we indeed see in Figure \ref{fig:btz-varyingm-from0101} that there are no qualitative changes in the Fisher information.  As $M$ decreases to small values, the curves shift leftward along the time axis, retaining their qualitative features. For large $M$, the curves converge, with essentially no discernible changes for $M > 1.5$.   We note from the bottom row (Dirichlet boundary condition) of Figure \ref{fig:btz-varyingm-from0101} that the maximal Fisher information increases for increasing $M$.  
While we have chosen to display $\Omega=T=0.1$ in this figure, similar behaviours are seen for other values of $\Omega=T$.

\begin{figure}
	\centering
	\includegraphics[width=1\linewidth]{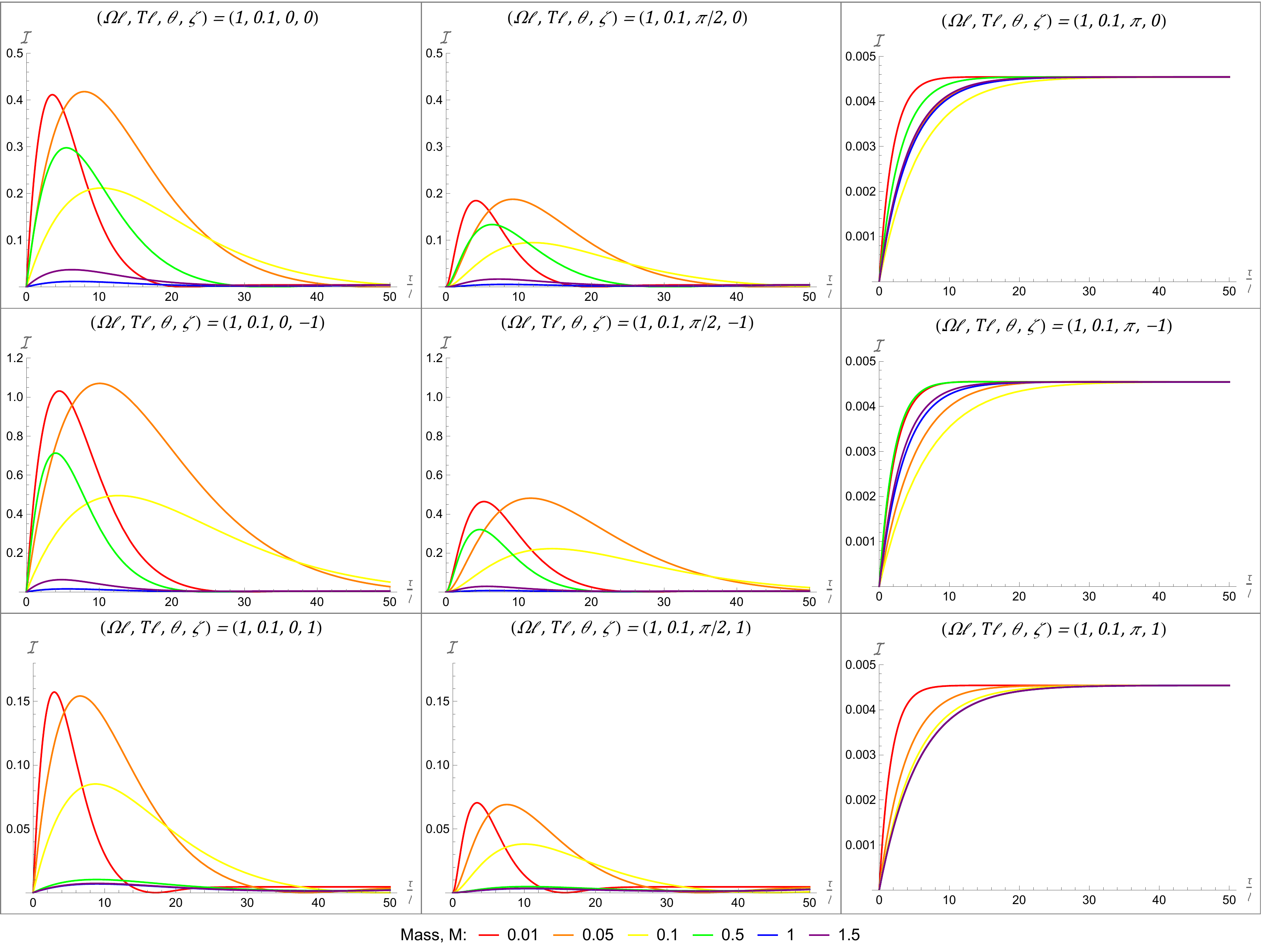}
	\caption[short caption]{We consider the effect of a variation in mass parameter, $M$, on the Fisher information given our `cold' set-up where $\Omega\ell=1$ and $T\ell=0.1$. This grid is set-up much like that displayed in Figure \ref{fig:btz-varyingm-from0101}, except that $\Omega \neq T$ here. This leads to non-monotonic behaviour when varying $M$. In particular, we see a smooth evolution from the red curve to the yellow one, but there is a change of direction as the Fisher information jumps back up from yellow to green. It then continues along its path down to the blue curve before jumping up once more to the purple curve. When considering the effect of the mass parameter more closely, we see that the Fisher information does indeed experience these two jumps. In fact, while these seem to be the only jumps, they are even more extreme than what is made visible here.
    Note that  the y-axis in the rightmost column (corresponding to $\theta=\pi$) has been rescaled by one to three orders of magnitude to highlight the non-zero asymptotic behaviour of the Fisher information for late time being fixed solely by $\Omega\ell$ and $T\ell$.
	}
	\label{fig:btz-varyingm-fromcold-grid}
\end{figure}

\begin{figure}
	\centering
	\includegraphics[width=0.9\linewidth]{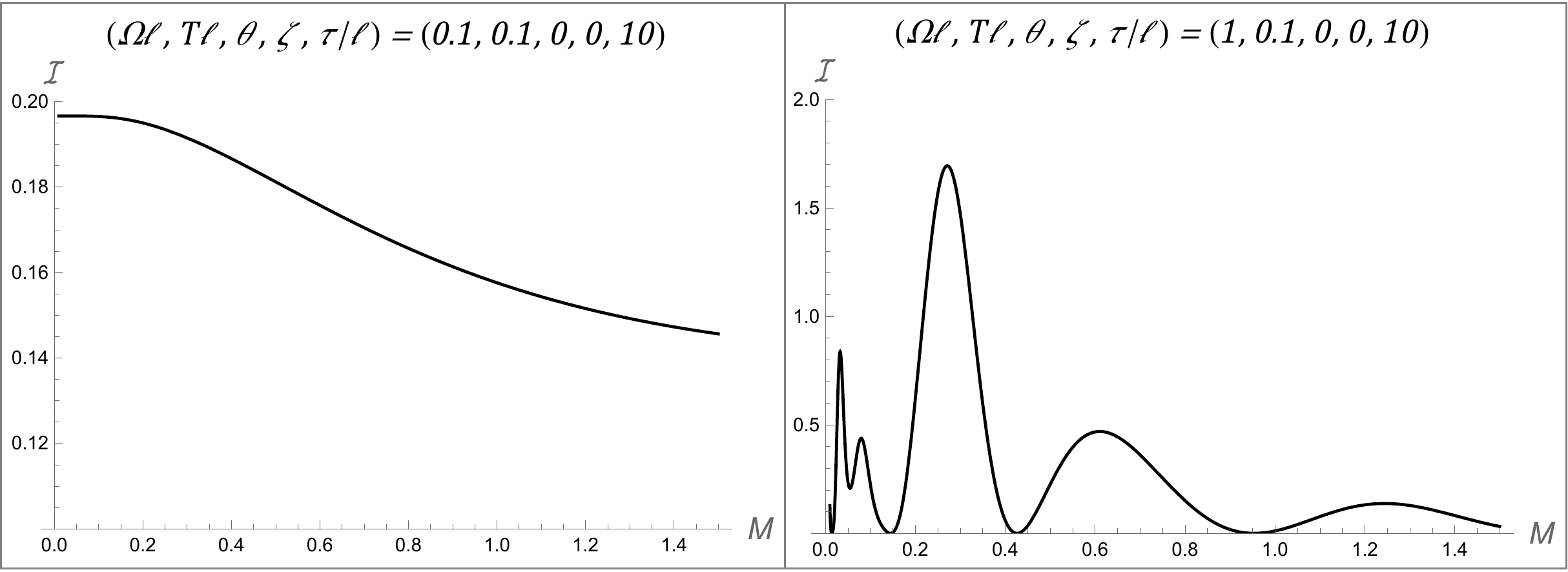}
	\caption[short caption]{ These plots show the effect  on the Fisher information of varying $M$ at a given time, chosen here to be $\tau/\ell=10$. Both plots are for the transparent boundary condition $(\zeta=0)$ with the initial state of the detector being $\theta=0$. On the left hand side we have the case in which increasing $M$ leads to a monotonic decrease in the Fisher information for $\Omega=T=0.1$. The right hand side shows   highly oscillatory behaviour present in the `cold' $\Omega\neq T$ case.}
	\label{fig:btz-varyingm-fixedtimecomparison}
\end{figure}

Turning next to situations where $\Omega \neq T$, the monotonic dependence on $M$ is lost  and we observe notably different behaviour, illustrated in Figure \ref{fig:btz-varyingm-fromcold-grid}.  For example in the `cold' case ($\Omega\ell=1$ and $T\ell=0.1$) we find that the Fisher information at any given time decreases from $M=0.01$ to $M=0.1$ before increasing and decreasing and then increasing again. We find that considering smaller increments of $M$ does not smooth out this behaviour but rather intensifies it. This is evident
from the right-hand diagram in   Figure \ref{fig:btz-varyingm-fixedtimecomparison}, which contrasts the effect of mass variation on the Fisher information for a fixed time (here $\tau/\ell=10$) between the $\Omega=T=0.1$ (on the left) and the `cold' $\Omega\neq T$ case. While this behaviour may not be evident here for the Dirichlet boundary condition (in the bottom row), it is present and visible when we examine the behaviour of the Fisher information  more closely while varying from $M=0.1$ to $M=0.5$.

 Figure \ref{fig:btz-varyingm-fixedtimecomparison} allows us to see how we might identify (or at least estimate) the mass of the black hole from the Fisher information. If, as in the left diagram, the parameters at play allow for a monotonic relation between the Fisher information and the mass, as in the left plot, then our job is easy: simply identify the value of the mass   associated with that particular value of the Fisher information. On the other hand, in the case of highly oscillatory behaviour (as in the right diagram), a slight change in the mass can have a large change on the Fisher information, which makes the peaks and troughs of the oscillations easier to identify. 

\begin{figure}[h!]
	\centering
	\includegraphics[width=\linewidth]{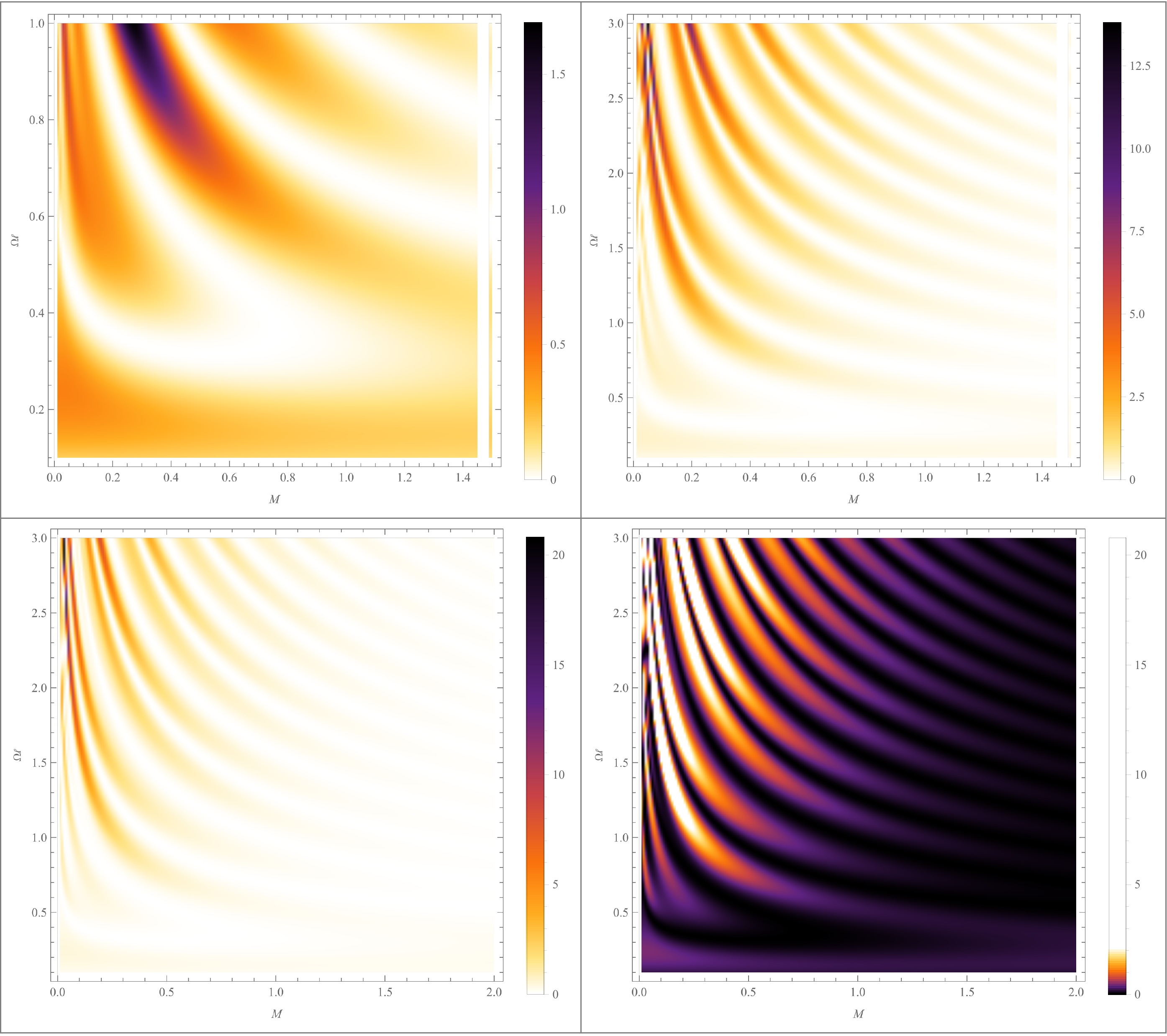}
	\caption[Short Caption]{  Depicted here are density plots of the Fisher information as a function of the energy gap $\Omega\ell$, along the vertical axes, and of the black hole mass $M$ along the horizontal axes. All other parameters are fixed to the same values as in Figure \ref{fig:btz-varyingm-fixedtimecomparison}, i.e., $T\ell=0.1$, $\zeta=0$, $\theta=0$, and $\tau/\ell=10$. In the top row, we cover the range in $\Omega\ell$ between the energy gaps used in Figure \ref{fig:btz-varyingm-fixedtimecomparison} on the left. We then extend this range to $\Omega\ell=3$ in the top right diagram. In the bottom row, we   plot the Fisher information for $0.01\leq\Omega\ell\leq3$ and $0.01\leq M\leq2$. The two bottom row plots depict the same data, but we have inverted and scaled the colour scheme in the plot on the right to highlight the behaviour for larger mass.  }
	\label{fig:btz-density}
\end{figure}

While Figure \ref{fig:btz-varyingm-fixedtimecomparison} presents us with two particular choices of energy gap, $\Omega\ell$, Figure \ref{fig:btz-density} depicts the Fisher information in a series of density plots where we vary both mass and energy gap. These allow us to see that this oscillatory   pattern is present for sufficiently large energy gaps (relative to the KMS temperature, i.e., sufficiently `cold' set-ups). Not only are there more oscillations for a fixed mass range in the colder set-ups, but the maximal amplitude of these Fisher information oscillations, occurring at smaller masses, is much greater. This suggests that   by choosing a large energy gap, we might significantly improve our thermal estimation procedure, that is, provide the appropriate small BTZ mass. 

It is worth noting that this oscillatory behaviour is most pronounced when the detector is initialized in the $\theta=0$ state. While it seems to be less extreme for $\theta=\pi$, there is still some `jumping around' present.  We observe similar results for the `hot' case, with $\Omega\ell=0.1$ and $T\ell=1$.

\section{Conclusion}\label{ch:conclusion}

We have carried out the first analysis of the Fisher information associated with the response of a UDW detector outside of a black hole.  We find   that Fisher information can act as an estimator for its  temperature and its mass.   
 
Taking specifically the $(2+1)$-dimensional black hole, we find the same qualitative behaviour for the Fisher information as a function of detector proper time as for its AdS-Rindler counterpart.
However quantitative differences between the two cases exist  that allow us to discriminate between the two. These are most noticeable for scenarios in which the detector gap $\Omega$ is not set equal to the temperature $T$.   

We also obtain the rather intriguing result that the Fisher information   exhibits an interesting dependence on the mass $M$ of the BTZ black hole.  For $\Omega=T$ the Fisher information monotonically depends on $M$.  However if $\Omega\neq T$ this monotonic behaviour is no longer present, and the Fisher information rapidly increases and decreases as $M$ increases from small values, exhibiting  highly oscillatory behaviour. These behaviours suggest that the Fisher information can function as an indicator of black hole mass.

Our results extend previous investigations in relativistic quantum metrology that seek to characterize different spacetimes based on their Fisher information. There remain a number of spacetimes that will likely offer interesting results subject to this sort of analysis. One such example would be black holes in $(3+1)$ dimensions, though the technical challenges there are more formidable.  Having said this, constant curvature topological black holes might provide a significant simplification \cite{deSouzaCampos:2020bnj,DeSouzaCampos:2022wsp}. Another interesting extension would be that of  analyzing the quantum Fisher information \cite{PetzGhinea2011} for these spacetimes. 

Perhaps the most straightforward extensions would be to other black holes in $(2+1)$ dimensions. These would include charged black holes, but also
rotating BTZ black holes,  which have been found to induce interesting entanglement behaviour arising from the quantum vacuum \cite{RHM2020}. It would be interesting to see if the thermal Fisher information is also sensitive to this  behaviour.  Likewise
an extension to the rotating 3-dimensional Gauss-Bonnet black hole 
\cite{Hennigar:2020drx} would be interesting, as the thermodynamics of this black hole is degenerate with the BTZ case.

\section*{Acknowledgements}

This work was supported in part by the Natural Sciences and Engineering Research Council of Canada (NSERC). E.P. acknowledges support from an NSERC Alexander Graham Bell CGS-M scholarship.

\bibliographystyle{./style_files/JHEP}
\bibliography{references}






\end{document}